# BEAM DYNAMICS OF THE SUPERCONDUCTING WIGGLER ON THE SSRF STORAGE RING *

ZHANG Qing-Lei(张庆磊)[1,2]  TIAN Shun-Qiang(田顺强)[1]  JIANG Bo-Cheng(姜伯承)[1]  XU Jie-Ping(许皆平)[1]
ZHAO Zhen-Tang(赵振堂)[1;1)]
[1] Shanghai Institute of Applied Physics, Chinese Academy of Sciences, Shanghai 201800, P.R. China
[2] University of Chinese Academy of Sciences, Beijing 100049, P.R. China

* Supported by National Natural Science Foundation of China (11105214)
1) E-mail: zhaozt@sinap.ac.cn

## *Abstract*

In the SSRF Phase-II beamline project, a Superconducting Wiggler (SW) will be installed in the electron storage ring. It may greatly impact on the beam dynamics due to the very high magnetic field. The emittance growth becomes a main problem, even after a well correction of the beam optics. A local achromatic lattice is studied, in order to combat the emittance growth and keep the good performance of the SSRF storage ring, as well as possible. Other effects of the SW are simulated and optimized as well, including the beta beating, the tune shift, the dynamic aperture, and the field error effects.

## INTRODUCTION

Shanghai Synchrotron Radiation Facility (SSRF) is a third generation light source with the beam energy of 3.5GeV [1-3]. It has been operated for users' experiments since 2009. There are 20 straight sections in the storage ring of SSRF, including 4 long straight sections (LSSs) and 16 standard straight sections (SSSs). Two of the LSSs have been occupied by the injection magnets and the RF cavities respectively, and eight of the SSSs have been installed with insertion devices (IDs). The SSRF Phase-II beamline project will be soon implemented in the near future, and more IDs will be equipped, including one set of SW to generate hard X-ray. The SW has much higher peak field than other IDs, which is a challenge to retain the storage ring performance. Main parameters of the existing IDs and this SW are listed in table 1.

Table 1: Main parameters the IDs in SSRF

| Name | Type | $\lambda_{ID}$ / mm | $L_{ID}$ / m | $B_{y,peak}$ / T |
|---|---|---|---|---|
| H08U | EPU | 100 | 4.3 | 0.60* |
| H09U58 | EPU | 58 | 4.9 | 0.68* |
| H09U148 | EPU | 148 | 4.7 | 0.67* |
| H13W | Wiggler | 140 | 1.4 | 1.94 |
| H14W | Wiggler | 80 | 1.6 | 1.20 |
| H15U | IVU | 25 | 2.0 | 0.94 |
| H17U | IVU | 25 | 2.0 | 0.94 |
| H18U | IVU | 25 | 1.6 | 1.00 |
| H19U1 | IVU | 20 | 1.6 | 0.84 |
| H19U2 | IVU | 20 | 1.6 | 0.84 |
| H03W | SW | 48 | 1.1 | 4.20~4.50 |

* For horizontal polarization mode

Since firstly installed to VEPP-3 [4], SW has been widely used in synchrotron light sources [5-12]. The magnetic field strength of SW is gradually raised thanks to the superconducting technology development. A peak field even up to 7.5T has been reached [9]. As the magnetic field strength increases, the critical energy of the photon emitted from SW is increased, and the influence of SW on the beam dynamics gets much stronger. Linear optics is distorted with SW, and compensation is achieved with quadrupoles in different ways, classified by local or global correction, and quadrupoles exited independently or in families. The non-linear effect is mainly in manner of dynamic aperture shrink, which would bring reduction of injection efficiency and beam lifetime. Magic finger has been used to eliminate the multipoles of SW in SPEAR [5] and SOLEIL [11], and high chromaticity is abated in CLS [8]. The horizontal emittance may increase after SW introduced. However, it can decrease if SW is located in low-dispersion or achromatic section [6, 9].

The peak magnetic field of the SW proposed for SSRF is 4.2T~4.5T, and in this paper we take 4.5T for our study. The SW may greatly impact on the beam dynamics of the storage ring. In order to retain the performance of the storage ring with SW, different schemes are tested. At first, a local optics correction is made by six quadrupoles adjacent to SW. Most of the beam parameters are restored, while the beam emittance increases by 39%, which will decrease the brightness of synchrotron radiation. Since reducing the dispersion at the SW straight section can help to depress oscillations of the electron caused by photon radiation in the SW, the beam emittance growth may be reduced. A local achromatic optics is considered then. This scheme is fulfilled with 20 quadrupoles in two cells adjacent to SW. The beam emittance growth is well suppressed. Dynamic aperture degrades severely due to non-cancellation of the nonlinear driving terms, so re-optimization with sextupoles [13] has been implemented. These results are presented in the following sections.

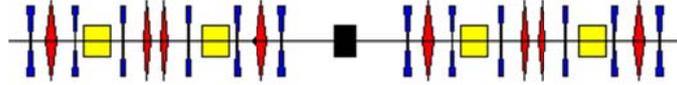

Figure 1: Layout of SW and adjacent two cells.

## TOOLS AND METHOD

Simulations are carried out with Accelerator Toolbox (AT) [14] working in MATLAB. The routines with intensive computation are written in C/C++, and compiled into MEX-files so as to be executable in MATLAB. As a result, AT can take the advantage of the efficiency in interactive modelling of MATLAB without losing its speed in time-consuming computation. AT is also open-ended, keeping progressing with efforts of people all over the world who are using and improving it [15].

The ID is considered to be a series of small dipole slices with hard edges. This model is checked with the current operation lattice, i.e. SSRF storage ring with the IDs in operation. As long as the slice number is sufficient, the simulation result of existing IDs with this model agrees well with the measurement. The slice number is a key point of this model. The simulation will be more credible as the slice number increases, however the computation time will also increase a lot. To figure out a reasonable slice number, the horizontal emittance is chosen as an object. The result is shown in figure 2, and the slice number is chosen to be 20 per magnetic field period, which is a good trade-off between the simulation accuracy and the computation time.

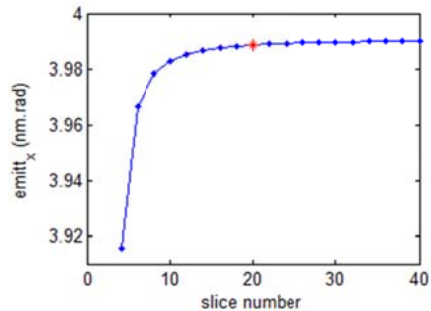

Figure 2: Horizontal emittance as an object of slice number per magnetic field period

The magnetic field is assumed to have a sinusoidal profile. The existing IDs and the SW are simulated in SSRF storage ring with AT. The emittance and energy spread of this simulation are consistent with calculation result from the formula in reference [16], as shown in table 2. The small variation between simulation and calculation is mainly from the approximation in the formula and the ID model.

Table 2: Simulation and calculation comparison

| | Emittance / nm·rad | | |
| --- | --- | --- | --- |
| | Bare lattice | Current lattice | Current lattice + SW |
| Simulation | 3.91 | 3.99 | 6.18 |
| Calculation | - | 4.00 | 5.98 |
| | Energy spread / $10^{-4}$ | | |
| | Bare lattice | Current lattice | Current lattice + SW |
| Simulation | 9.83 | 9.79 | 10.7 |
| Calculation | - | 9.76 | 11.7 |

## OPTICS DISTORTION AND CORRECTION

The SW significantly disturbs the linear optics of the SSRF storage ring. The vertical tune shift is 0.016, and the maximum vertical beta beating is above 10%, while the horizontal variations are not distinct. Figure 3 plots the beta beating in the whole ring, where the RMS beta beatings are 0.0013% and 7.4% in the horizontal and vertical plane, respectively. The SW distorts the beam dynamics much more in the vertical plane than the horizontal plane, because the SW, as a planar ID, has vertical magnetic field only.

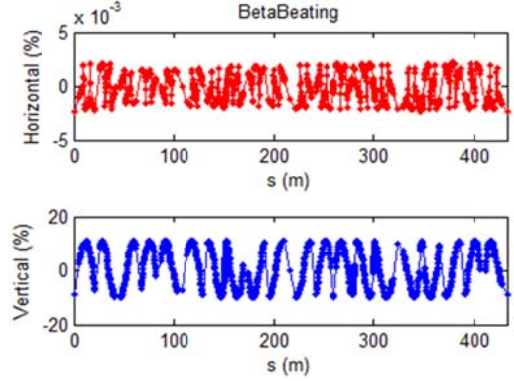

Figure 3: Global beta beating of SW

In the current settings of the SSRF storage ring, the RMS beta beating of the bare lattice is around 10% without optics correction, and it can be efficiently reduced to about 1% by the individually excited quadrupoles [17]. This fact provides great confidence to compensate the distortions from SW. In the interest of minimizing the adjustment, a local correction scheme is applied. Only six quadrupoles adjacent to SW are used to correct the optics, and the minimum global RMS beta beatings are the objects. After well correction, the RMS beta beatings are reduced to about 0.5% in the two transverse plans. However, the emittance growth is still large as 39%, which is reduced from 55% before correction. The variation of energy spread is negligible.

## LOCAL ACHROMATIC SCHEME

In order to reduce the emittance growth, a local achromatic lattice is considered, i.e. the dispersion is impressed to zero from 0.105m in the standard straight where SW locates, while the dispersion and the beta functions out of this section are kept the same as the current settings in order not to change the parameters of other source points. There are two constraints to achieve this scheme[18], so at least two quadrupoles before the bending magnet are needed. To match the beta function, another four quadrupoles are needed. Actually, all the quadrupoles in the adjacent two cells are involved so as to get a solution with reasonable strength.

Figure 4 plots the new achromatic optics in the two cells, comparing with the current optics. There are some variations in the involved two cells, but the optics is kept perfectly consistent besides the two cells.

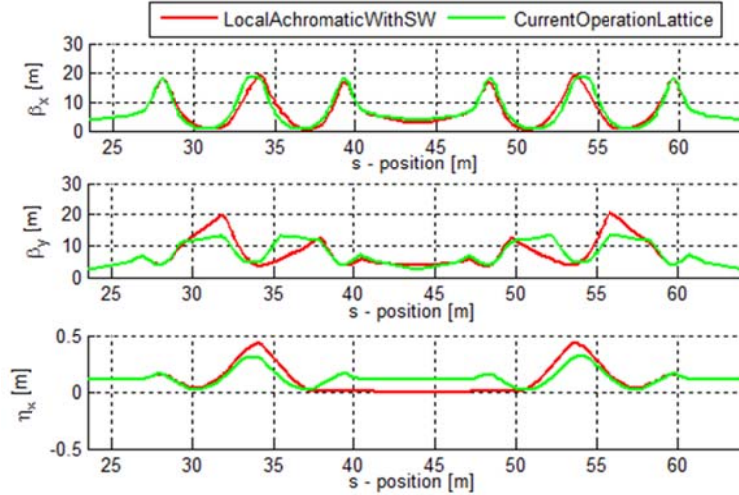

Figure 4: Optics of the two cells adjacent SW

The natural emittance is reduced to 4.09nm·rad by the more strong radiation damping of SW in this scheme. There is only an emittance growth of about 3% from the current lattice without SW to the new local achromatic lattice with SW, which is acceptable. Table 3 summarizes the beam parameters of the new local achromatic lattice and the current lattice, where the IDs in operation are included in both lattices.

Table 3: Beam parameters of different scheme

|  | **Current lattice** | **New lattice** |
| --- | --- | --- |
| Emittance / nm·rad | 3.99 | 4.09 |
| Energy spread / $10^{-4}$ | 9.8 | 10.7 |
| Energy loss per turn / MeV | 1.52 | 1.69 |
| Tune (H, V) | 22.22, 11.30 | 22.27, 11.30 |
| Natural Chromaticity (H, V) | -55.7, -18.0 | -55.5, -18.2 |
| Mom. comp. factor / $10^{-4}$ | 4.3 | 4.5 |
| Max beta (H, V) / m | 25.4, 15.9 | 25.4, 20.6 |
| Max eta / m | 0.321 | 0.438 |
| Energy acceptance | 3.75% | 3.49% |
| Bunch length / ps | 12.1 | 13.6 |

In the achromatic scheme, the major trouble of emittance increase has been solved, nevertheless the dynamic aperture degrades severely, as shown in figure 5 (blue line). Small dynamic aperture will reduce the injection efficiency or the beam lifetime. To obtain a sufficient dynamic aperture, re-optimization with sextupoles is necessary. The dynamic aperture enlarges after carefully adjustment of the sextupole strength, which is shown in figure 5 (green line). The horizontal dynamic aperture is restored, which means undegraded injection efficiency. The vertical dynamic aperture is optimized to exceed the physical aperture.

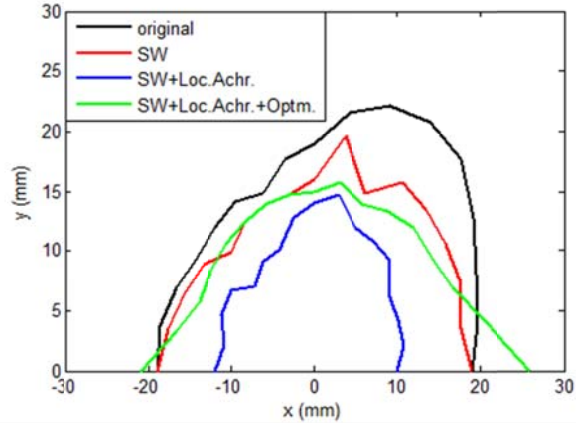

Figure 5: Dynamic aperture of different scheme with 1000 turns tracking by AT.

## FIELD ERROR ANALYSIS

Imperfection is unavoidable in ID manufacturing process, which will introduce errors in magnetic field. Errors will bring variation in performance, and error analysis is essential to estimate the robustness of the lattice.

Since the ID is an array of magnet blocks, the errors will be block-dependant. So a random error array corresponding to the block array is generated. Nevertheless this array is too rough to simulate a continuous distribution. Hence we reform this array using interpolation algorithm.

To simulate errors, we need to define the type and strength of them. Here we consider the errors of dipole and quadrupole, which are the main concern. The integral dipole error is investigated from 0 to $\pm 1000$Gauss·cm, while integral quadrupole error varies from 0 to $\pm 1000$Gauss.

On the other hand, there are different distributions that satisfy a certain integral dipole or quadrupole error, and different distribution will lead to different impact to beam dynamics. In practice, 100 random distributions for each integral error are generated, and all their impacts on beam dynamics are investigated. The variations of tune, chromaticity, energy spread and horizontal emittance are observed, and the result is summarized in figure 6. The relationship between these parameters and errors is reasonable, and the variations of these parameters are tolerable.

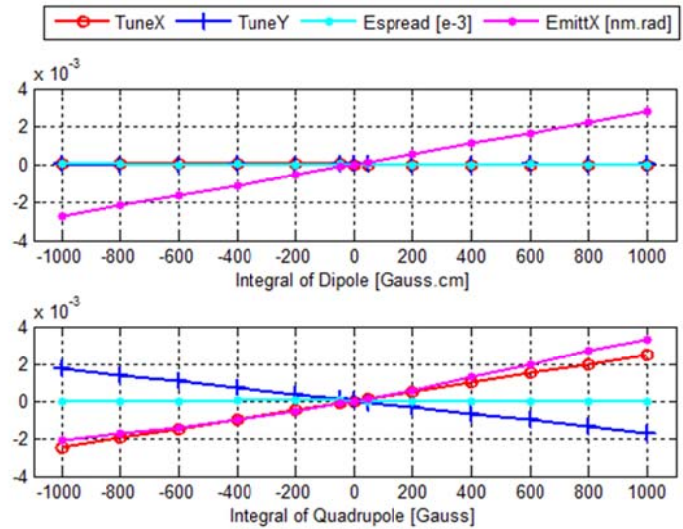

Figure 6: The variations of main parameters with errors.

# CONCLUSION

The effects of a 4.5T SW on beam dynamics in SSRF has been studied. Significant impact on optics has been found, and the emittance increases by 55%, which will greatly reduce the brightness of synchrotron radiation. Global linear optics is restored with quadrupole correction, and the emittance growth is 39% then. For further optimization, a local achromatic lattice is considered so as to enhance the damping effect of the SW. The emittance growth is reduced to 3% without global optics distortion in this scheme, which is tolerable for the machine. However, the dynamic aperture degrades severely, which is a problem to the lifetime and injection efficiency. Optimization with sextupoles has been implemented, and the dynamic aperture is enlarged to retain the lifetime and injection efficiency. Error analysis has also been studied. The variations of main parameters are tolerable with integral dipole and quadrupole errors up to $\pm 1000$Gauss·cm and $\pm 1000$Gauss, respectively.